\documentclass[twocolumn]{aastex631}

\usepackage{amsmath}	
\newcommand{\synax}{\texttt{synax}}
\newcommand{\hp}{\texttt{healpix}}
\newcommand{\hpy}{\texttt{healpy}}
\usepackage{newtxtext,newtxmath}
\usepackage{fontawesome}

\submitjournal{ApJS}

\received{XXX}
\revised{YYY}
\accepted{ZZZ}

\shorttitle{\synax\ Synchrotron Simulation}

\shortauthors{Diao et al.}

\begin{document}

\title{\synax: A Differentiable and GPU-accelerated Synchrotron Simulation Package}

\author[0000-0001-7301-2318]{Kangning Diao}
\affiliation{Department of Astronomy, Tsinghua University, Beijing, 100084, China}
\affiliation{Berkeley Center for Cosmological Physics, University of California, Berkeley, CA 94720, United States}

\author[0000-0002-0309-9750]{Zack Li}
\affiliation{Berkeley Center for Cosmological Physics, University of California, Berkeley, CA 94720, United States}

\author[0000-0001-9578-6111]{Richard D.P. Grumitt}
\affiliation{Department of Astronomy, Tsinghua University, Beijing, 100084, China}

\author[0000-0002-1301-3893]{Yi Mao}
\affiliation{Department of Astronomy, Tsinghua University, Beijing, 100084, China}

\correspondingauthor{Kangning Diao, Yi Mao}
\email{dkn20@mails.tsinghua.edu.cn (KND), ymao@tsinghua.edu.cn (YM)}



\begin{abstract}

We introduce \synax\ \href{https://github.com/dkn16/Synax}{\faGithub}, a novel library for automatically differentiable simulation of Galactic synchrotron emission. Built on the JAX framework, \synax\ leverages JAX's capabilities, including batch acceleration, just-in-time compilation, and hardware-specific optimizations (CPU, GPU, TPU). Crucially, \synax\ uses JAX's automatic differentiation (AD) mechanism, enabling precise computation of {analytical} derivatives with respect to any model parameters. This facilitates powerful inference algorithms, such as Hamiltonian Monte Carlo (HMC) and gradient-based optimization, which enables inference over models that would otherwise be computationally prohibitive. In its initial release, \synax\ supports synchrotron intensity and polarization calculations down to GHz frequencies, alongside several models of the Galactic magnetic field (GMF), cosmic ray (CR) spectra, and thermal electron density fields. {When running \synax\ on the CPU we obtain identical performance to \texttt{hammurabi}, a start-of-the-art synchrotron simulation package, while on the GPU \synax\ brings a twenty-fold enhancement in efficiency. We further demonstrate the potential of AD in enabling full posterior inference using gradient-based inference algorithms. Using \synax\ with HMC to perform inference over a four-parameter test model, we attain a two-fold improvement compared to standard random walk Metropolis-Hastings (RWMH). When applied to a more complex 16-parameter model, HMC is still able to obtain accurate posterior expectations while RWMH fails to converge. We also showcase the application of \synax\ to optimizing the GMF based on the Haslam 408 MHz map, achieving residuals with a standard deviation below 1 K.}

\end{abstract}



\section{Introduction}

Galactic synchrotron emission dominates the low-frequency radio sky, spanning frequencies from MHz to GHz, and obscures cosmological signals, including those from the cosmic microwave background (CMB; \citealt{S-PASS,C-BASS,Quijote}), the 21 cm line \citep{2020PASP..132f2001L}, and other line intensity measurements \citep[e.g.][]{2014ApJ...785...72G,2023PhRvD.107l3504M}. This emission originates from the interaction between cosmic ray (CR) electrons and the Galactic magnetic field (GMF), with polarization altered as a result of Faraday rotation in the presence of the GMF and thermal electrons \citep{1986rpa..book.....R}. Consequently, while mitigating synchrotron emission is essential for extracting cosmological signals, the emission itself also serves as a probe for the structure of the Galactic interstellar medium (ISM) and CR transport processes \citep[e.g.][]{2024ApJ...970...95U}.

During the past decades, several simulation packages have been developed to model Galactic synchrotron emission, including \texttt{hammurabi} \citep{hammurabi,hammurabiX} and \texttt{ULSA} \citep{ULSA}. \texttt{hammurabi}, when combined with models of the GMF \citep{WMAP1yr,JF12}, CRs \citep{WMAP1yr}, and thermal electrons \citep{YMW16}, has successfully reproduced the large-scale components of the Galactic disc but has struggled to capture finer structures. \texttt{hammurabi} has been employed in simulation-based inference to study magnetic fields in astrophysical contexts, ranging from supernova remnants \citep{2017ApJ...849L..22W} to the entire Galaxy \citep{JF12,2016A&A...596A.104P}. \texttt{ULSA} extends these capabilities by incorporating free-free emission and absorption, enabling precise simulations down to 1 MHz. Utilizing \texttt{ULSA}, Markov chain Monte Carlo (MCMC) methods have been applied to infer the local three-dimensional distribution of thermal electrons \citep{2022ApJ...940..180C}.

However, synchrotron simulations remain computationally demanding, requiring integration along numerous sightlines with fine resolution. These operations can be highly parallelized because of the independence between different sightlines. Recently, \texttt{JAX} \citep{deepmind2020jax} has emerged as a multiplatform computing framework, supporting CPUs, GPUs, TPUs, and offering automatic differentiation (AD). GPUs, designed for parallel computation, are particularly well suited for synchrotron simulations. Moreover, AD provides access to gradients, allowing more powerful sampling and optimization algorithms, which are crucial to conducting inference on more complex models \citep{gelman1997weak, beskos2013hmc}.

In this work, we present \synax\ \citep{synax_zenodo}, a synchrotron simulation package powered by \texttt{JAX}. The code is implemented in a parallelized manner, significantly accelerating sampling and optimization, enabling inference over high-dimensional models at the field-level. {We investigate the performance of \synax\ as applied to inference tasks, starting with posterior inference over simple analytical models of Galactic fields with only a handful of parameters, and then assess the optimization performance on a high-dimensional field-level analysis.}

This paper is organized as follows: In Section \ref{sec:method}, we describe the theory and formalism for computing synchrotron emission. {In Section \ref{sec:inference} we demonstrate the capabilities of \synax\ in examples using fast, gradient-based inference algorithms. In Section \ref{sec:discussion} we discuss the error budget and performance of \synax\ with respect to the choice of integration points, and demonstrate sampling over a 16-parameter model and optimization of the GMF based on the 408 MHz Haslam map \citep{Haslam}. We summarize our findings in Section \ref{sec:conclusion}.}

\section{Method}
\label{sec:method}
In this section, we follow \citet{1986rpa..book.....R} and describe the physical processes and the numerical techniques used to compute the synchrotron emission maps.

\subsection{Synchrotron Intensity}
Galactic synchrotron emission is caused by the spiraling of relativistic charges in the GMF. We assume that CR electrons generated by supernova explosions dominate the population of Galactic relativistic charges, and that these electrons subsequently experience shock acceleration. We further simplify the problem by assuming an isotropic velocity distribution for CR electrons, which has been measured with high accuracy to be the case at our location in the Galaxy \citep{Yan2008}. A full CR spectrum treatment will be addressed in future work, {which will allow us to forward model more accurate synchrotron spectra beyond simple power-laws and perform inference over them with gradient-based sampling}.

The CR electron spectrum is typically modeled as a power law with a spectral index \(\alpha\). This widely used simplification is supported by the theory of shock acceleration \citep{Drury1983} and has been broadly adopted in previous work \citep{ULSA,hammurabi,hammurabiX}. This assumption has also been successfully applied to CMB foreground subtraction \citep[][]{2020A&A...641A...4P,Eriksen_2008,planckcomponent} and corroborated by measurements of the CR electron spectrum on Earth \citep[e.g.][]{2011PhRvL.106t1101A,2013FrPhy...8..748G}. While both observations and advanced simulations have shown that the power-law assumption is inadequate for representing the entire spectrum \citep[e.g.][]{Strong2007}, accurately accounting for the full spectrum is significantly more time-consuming. Therefore, we adopt the assumption of a power-law spectrum for simplicity in this work, {in order to balance accuracy and computational efficiency}, leaving a more detailed treatment for future work.

Given the physical processes and assumptions outlined above, the specific intensity of synchrotron emission $I(\nu,\hat{\mathbf{n}})$ at {a given} frequency $\nu$ and along the line-of-sight (LOS) direction $\hat{\mathbf{n}}$ is given by 
\begin{equation}
    I(\nu,\hat{\mathbf{n}}) = \int_{0}^{\infty} j_{\rm I}(\nu,r'\hat{\mathbf{n}}+\mathbf{r}_{\rm obs}) dr' \,,
\end{equation}
where $r'$ is the distance of a field point to the observer on Earth and $\mathbf{r}_{\rm obs} = (-8.3,0.0,0.006)$ kpc is the distance vector from the Galactic center to the Earth. Here, $j_{\rm I}(\nu,\mathbf{r})$ is the emissivity of the synchrotron specific intensity, where $\mathbf{r}=r'\hat{\mathbf{n}}+\mathbf{r}_{\rm obs}$ is the distance vector from the Galactic center to the field point. The emissivity $j_{\rm I}$ is given by \citep{1959ApJ...130..241W} 
\begin{equation}
\begin{aligned}
    j_{\rm I}(\nu,\mathbf{r}) &=\frac{\sqrt{3}q_e^3B_{\rm trans}(\mathbf{r})N_0(\mathbf{r})}{8\pi m_e c^2}\left ( \frac{4\pi \nu m_e c}{3 q_e B_{\rm trans}(\mathbf{r})}\right )^{(1-\alpha)/2}\\
    &\times\frac{2^{(\alpha+1)/2}}{\alpha+1}\Gamma\left ( \frac{\alpha}{4}-\frac{1}{12}\right)\Gamma\left ( \frac{\alpha}{4}+\frac{19}{12}\right).
    \label{eqn:ji}
\end{aligned}
\end{equation}
The amplitude of the GMF transverse to the LOS direction is given by $B_{\rm trans}$, $q_e$ is the electron charge, $m_e$ is the electron mass and $c$ is the speed of light. We assume that high-energy CR electrons follow a power law $N(\gamma,\mathbf{r})d\gamma = N_0(\mathbf{r})\gamma^{-\alpha}d\gamma$, where $N_0(\mathbf{r})$ is the normalization factor and $\gamma$ is the Lorentz factor.

\subsection{Synchrotron Polarization}

For synchrotron polarization, we focus on the specific intensity of polarization $P$. We further define the parameters $Q$ and $U$, which describe the specific intensity of linear polarization, through $P=Q+iU$. The quantities $\{I,Q,U\}$ are related with the first three Stokes parameters $\{S_0,S_1,S_2\}$ \citep{perrin1942polarization} by
\begin{equation}
        S_0 = \int I d \Omega, \qquad
        S_1 = \int Q d \Omega, \qquad
        S_2 = \int U d \Omega,
\end{equation}
where $\Omega$ is the solid angle. 
The {specific intensity of} polarization $P$ is given by 
\begin{equation}
    P(\nu,\hat{\mathbf{n}}) = \int_{0}^{\infty} j_{\rm P}(\nu,r'\hat{\mathbf{n}}+\mathbf{r}_{\rm obs})e^{2i\chi(r'\hat{\mathbf{n}}+\mathbf{r}_{\rm obs})} dr',
\end{equation}
where $j_{\rm P}$ is the synchrotron polarized emissivity, given by \citep{1959ApJ...130..241W,1966MNRAS.133...67B} 
\begin{equation}
\begin{aligned}
    j_{\rm P}(\nu,\mathbf{r}) &=\frac{\sqrt{3}q_e^3B_{\rm trans}(\mathbf{r})N_0(\mathbf{r})}{8\pi m_e c^2}\left ( \frac{4\pi \nu m_e c}{3 q_e B_{\rm trans}(\mathbf{r})}\right )^{(1-\alpha)/2}\\
    &\times 2^{(\alpha-3)/2}\Gamma\left ( \frac{\alpha}{4}-\frac{1}{12}\right)\Gamma\left ( \frac{\alpha}{4}+\frac{7}{12}\right),
    \label{eqn:jp}
\end{aligned}
\end{equation}
and $\chi(\mathbf{r})$ is the observed polarization angle of the polarized {signal} emitted at $\mathbf{r}$. 

Synchrotron radiation undergoes Faraday rotation as it travels through a magnetized plasma, causing the polarization angle to vary with frequency. The polarization angle $\chi$ can be modeled as \citep[e.g.][]{1986rpa..book.....R}
\begin{equation}
    \chi(\mathbf{r}) = {\rm RM} \cdot\lambda^2+\chi_0(\mathbf{r}),
\end{equation}
where  $\lambda$ is the wavelength and the rotation measure (RM) quantifies the linear rate of change of the angle $\chi$. RM, which depends on $\mathbf{r}=r'\hat{\mathbf{n}}+\mathbf{r}_{\rm obs}$, can be calculated via
\begin{equation}
    {\rm RM} = \frac{q_e^3}{2\pi m_e^2c^4}\int_0^{r'} n_e(r''\hat{\mathbf{n}}+\mathbf{r}_{\rm obs}) B_{\rm LOS}(r''\hat{\mathbf{n}}+\mathbf{r}_{\rm obs}) dr'',
    \label{eqn:rm}
\end{equation}
where $n_e$ is the electron density and $B_{\rm LOS} = \Vert \mathbf{B}\cdot \hat{\mathbf{n}}\Vert $ is the amplitude of the LOS component of the magnetic field. 

The intrinsic polarization angle $\chi_0$ is defined as \citep[e.g.][]{hammurabiX} 
\begin{equation}
    \tan (\chi_0) = \frac{B_z \cos(b) - B_x\cos(l)\sin(b) - B_y\sin(l)\sin(b)}{B_y\sin(l)-B_x\cos(l)}\,.
    \label{eqn:chi}
\end{equation}
Here $(B_x,B_y,B_z)$ are the $x$, $y$ and $z$-components of $\mathbf{B}(\mathbf{r})$ field respectively, where the directions of the axes are defined in Galactic coordinates, and $l$ is the Galactic longitude and $b$ is the Galactic latitude for the LOS as seen by the observer on Earth. 

\subsection{Integration}

The input fields required when simulating synchrotron emission with \synax\ are:
\begin{enumerate}
    \item 3D GMF $\mathbf{B}(\mathbf{r})$,
    \item 3D thermal electron density distribution $n_e(\mathbf{r})$,
    \item Normalizing factor of the CR electron distribution $N_0(\mathbf{r})$,
    \item Power-law index of the CR energy spectrum $\alpha$.
\end{enumerate}
We follow \citet{WMAP1yr} and set the power law index $\alpha=3$ throughout this paper unless otherwise noted.
\begin{figure}
    \centering
    \includegraphics[width=\linewidth]{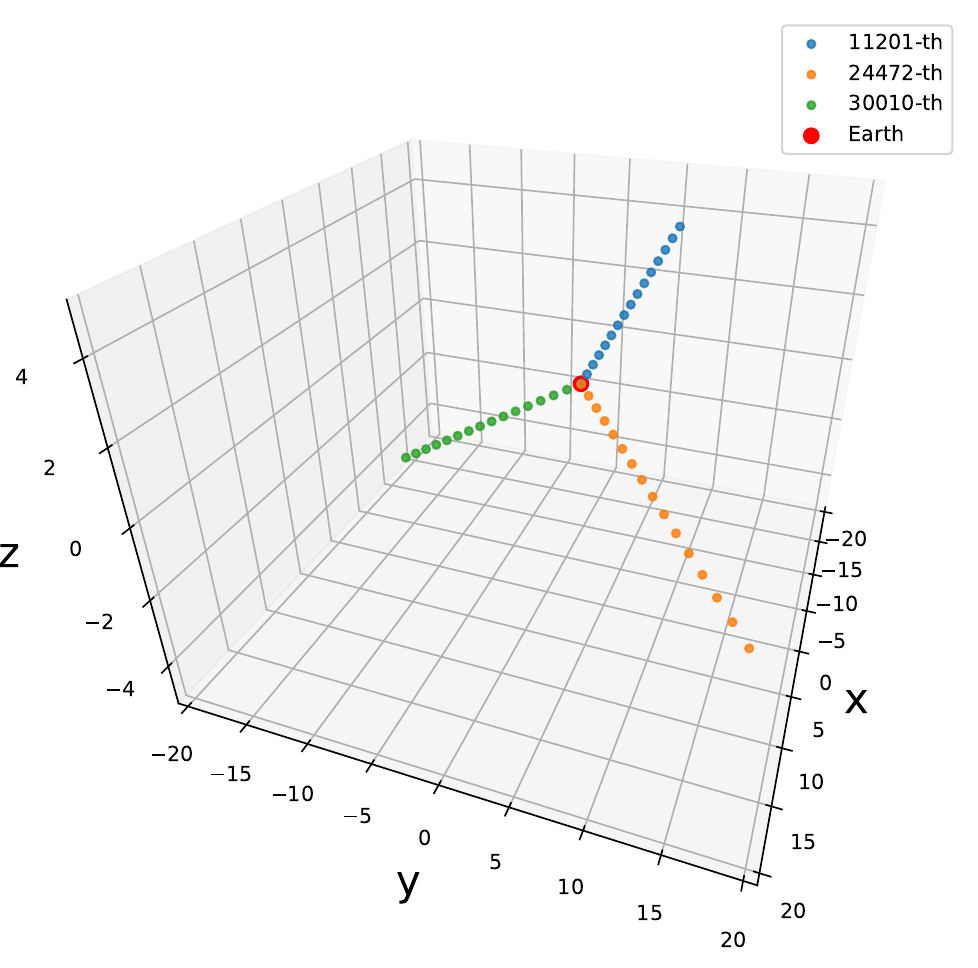}
    \caption{An illustration of the integration points along several sightlines with the number of integration points $N_{\rm int} = 16$ in a box of length $(40,40,10)$ kpc along the $x$, $y$, $z$ directions respectively. The legend shows the index of the sightline for an NSIDE = 64 \hp\ map, while the large red dot represents the observer on Earth.}
    \label{fig:int_points}
\end{figure}

For the fields $\{\mathbf{B}(\mathbf{r}), n_e(\mathbf{r}), N_0(\mathbf{r})\}$, \synax\ accepts a callable field generator function as input. This function can either be an analytical model or an interpolation function operating over {regular} 3D grids (hereafter 3D grids). To ensure consistency between 3D grids and analytical functions, we assume that all fields are confined within a 3D box of dimensions $({l_x, l_y, l_z})$ centered at the Galactic center, with contributions from outside the box being considered negligible.

We begin by generating the coordinates for all integration points. For the $i$-th sightline in the \hp\ map, the coordinates $\{l_i, b_i\}$ are obtained using \hpy. The intersection point between the sightline and the boundary of the 3D box is then calculated. The entire LOS from the observer to the boundary is uniformly divided into $N_{\rm int}$ segments with the segment length $\Delta r_i$, with the midpoint coordinates $\{\mathbf{r}_{i,n}; n=1,2,\ldots,N_{\rm int}\}$ being computed for each segment. The field values $\{\mathbf{B}(\mathbf{r}_{i, n}), n_e(\mathbf{r}_{i,n}), N_0(\mathbf{r}_{i, n})\}$ are then evaluated at each of the midpoints. Figure \ref{fig:int_points}
provides an illustration of the integration points along several sightlines. If an analytical function is provided for a field, its value at $\mathbf{r}_{i,n}$ is calculated analytically. Otherwise, we determine the field values using 3D linear interpolation, implemented with  \texttt{Interpax}\footnote{\url{https://interpax.readthedocs.io/en/latest/}}.

With these field values, $\chi$, $j_{\rm I}$, and $j_{\rm P}$ can be computed at each $\mathbf{r}_{i,n}$ using Equations~(\ref{eqn:ji})-(\ref{eqn:chi}). Discretizing along the sightline and multiplying by the segment length $\Delta r_i$ yields the specific intensities,
\begin{equation}
\begin{aligned}
    I(\nu,\hat{\mathbf{n}}_i) &= \sum_{n=1}^{N_{\rm int}}j_{\rm I}(\nu,\mathbf{r}_{i,n})\Delta r_i,\\
    Q(\nu,\hat{\mathbf{n}}_i) &= \sum_{n=1}^{N_{\rm int}}j_{\rm P}(\nu,\mathbf{r}_{i,n})\cos(2\chi(\mathbf{r}_{i,n}))\Delta r_i.\\
    U(\nu,\hat{\mathbf{n}}_i) &= \sum_{n=1}^{N_{\rm int}}j_{\rm P}(\nu,\mathbf{r}_{i,n})\sin(2\chi(\mathbf{r}_{i,n}))\Delta r_i,
    \label{eqn:int}
\end{aligned}
\end{equation}


\subsection{Validation and Performance on a Toy Model}
\begin{figure*}
\begin{center}
\begin{tabular}{c}
\includegraphics[width=0.85\linewidth]{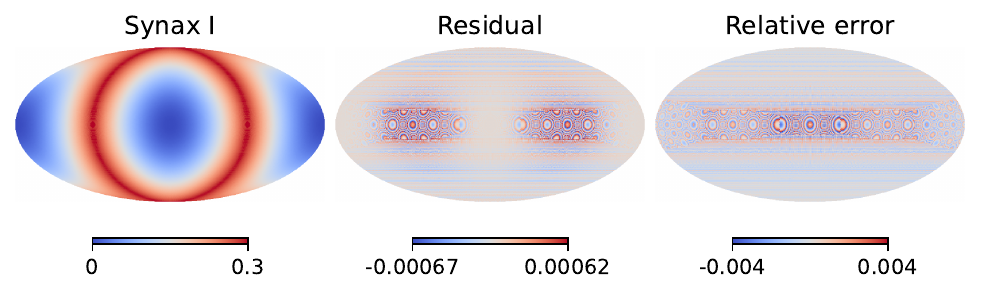}\\
\includegraphics[width=0.85\linewidth]{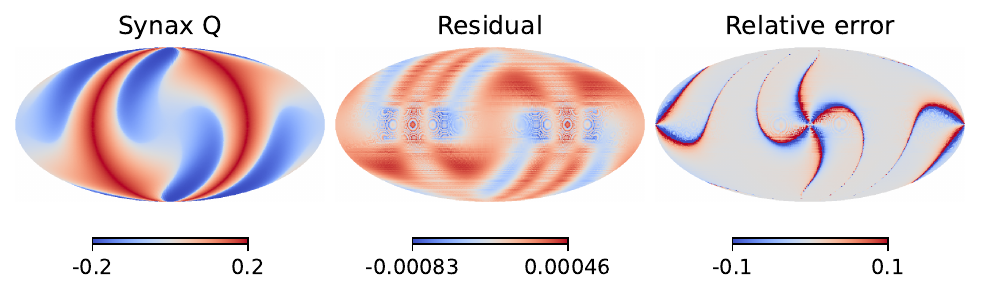}\\
\includegraphics[width=0.85\linewidth]{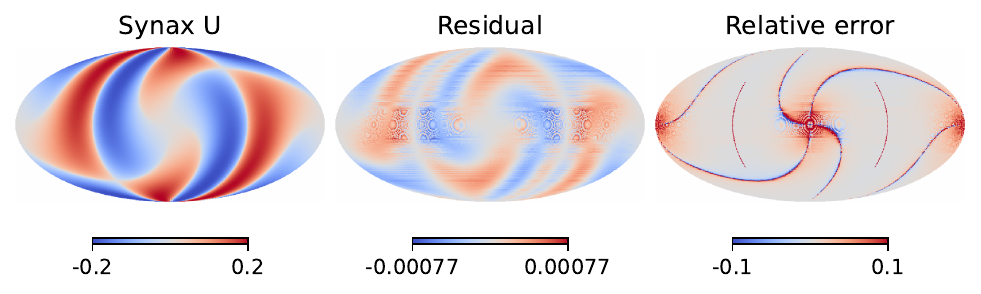}
\end{tabular}
\end{center}
\caption{Accuracy test with homogeneous $\{\mathbf{B}(\mathbf{r}),n_e(\mathbf{r}),N_0(\mathbf{r})\}$ fields with callable field generators. The left panel shows the synchrotron $\{I,Q,U\}$ maps generated by \synax, the middle panel shows the residual between \synax\ and theoretical value, and the right panel shows the relative error between \synax\ and the theoretical value. Unless mentioned otherwise, all \hp\ maps in this work are in units of Kelvin.}
\label{Fig:acc_func}
\end{figure*}

We first test our model on a simple example \citep{hammurabiX} at $\nu = 2.4$ GHz. In this model, the fields are homogeneous, with $\mathbf{B}(\mathbf{r})=(6\times10^{-6} {\rm gauss},0,0)$, $n_e(\mathbf{r}) = 0.01\,{\rm cm^{-3}}$, and $N_0(\mathbf{r}) = 4.01 \times 10^{-5}{\rm cm^{-3}}$ within a spherical region of the radius $R_0 = 4$ kpc {around} the Earth, and zero outside (note that these fields are not spherically symmetric with respect to the Galactic center). In this case, the emission intensity before Faraday rotation $\{I,Q_0,U_0\}$ becomes,
\begin{equation}
    \begin{aligned}
        I &= j_{\rm I}R_0,\\
        Q_0 &= j_{\rm P}R_0\cos(2\chi_0),\\
        U_0 &= j_{\rm P}R_0\sin(2\chi_0),
    \end{aligned}
\end{equation}
since $j_{\rm I},j_{\rm P},\chi_0$ are constants along a {given} sightline. {The polarized signal after Faraday rotation is given by}
\begin{equation}
    Q+iU = (Q_0+iU_0)\int_0^{R_0}\frac{1}{R_0}e^{2i\lambda^2r'n_eB_{\rm LOS}q_e^3/2\pi m_e^2c^4}dr'.
\end{equation}
{Given these expressions}, $\{I,Q,U\}$ maps can be calculated analytically. 

We simulate the $\{I,Q,U\}$ maps with \synax, {setting ${\rm NSIDE} = 64$, $N_{\rm int}=1024$, and the length of the box to $(40,40,10)$ kpc}. {The input fields are given by} analytical functions and the results are shown in Figure \ref{Fig:acc_func}. The typical scale of absolute residuals is below 1\% of the signal scale {and for every doubling of the integration points, the error is halved.}\footnote{The range of integration points we tested is from 512 to 4096.} The relative error, defined by $\epsilon_{\rm rel} =2(X_{\rm sim}-X_{\rm ana})/(X_{\rm sim}+X_{\rm ana})$ where $X_{\rm sim}$ is the simulated map and $X_{\rm ana}$ is the analytical map, is also mostly at the percentage level, except for sightlines with field values very close to zero. We note that the residuals are all below 1 mK, suggesting that \synax\ with {analytical function inputs} has no significant bias {in the context of} mK noise level observations.


Our code is based on \texttt{JAX} \citep{deepmind2020jax} and can run on multiple platforms including CPU, GPU and TPU. {The default precision is double precision, which is also the default precision for all examples in this work}. {For the above example}, \synax\ uses {$\lesssim16$} ms to generate the emission map on an NVidia Tesla A100 and {$\sim810$ ms with {128} threads on an AMD EPYC 7763 CPU}, after just-in-time (JIT) compilation. For comparison, it takes {$\sim800$} ms for the MPI-parallelized CPU-based code \texttt{hammurabi} \citep{hammurabiX} to run with {128} threads on an AMD EPYC 7763 CPU {with the same number of integration points}. \synax\ obtains the same level of accuracy as \texttt{hammurabi}\footnote{{Here we use the python wrapper of \texttt{hammurabi}, namely \texttt{hampyx}. The number of integration points is adjusted to have the same integration within the sphere for the evaluation of accuracy.}}, with the standard deviation of residuals for the $\{I,Q,U\}$ maps with \synax\ being $\{1.6\times10^{-4},2.0\times10^{-4},2.0\times10^{-4}\}$, while with \texttt{hammurabi} the standard deviation of residuals is $\{1.6\times10^{-5},5.6\times10^{-4},6.0\times10^{-4}\}$ respectively. 
\synax\ gains {$\sim40$ times}\footnote{{The $\sim 40$ times improvement is estimated with the economic consideration that an NVidia Tesla A100 and an AMD EPYC 7763 CPU have similar prices at the moment of preparing this manuscript. However, the theoretical computability is approximately 10 (2.5) TFLOPs for NVidia Tesla A100 (AMD EPYC 7763). If the performance difference between GPU and CPU is corrected, \synax\ gains $\sim 10$ times improvement due to GPU acceleration.}} improvement {on {the} GPU and performs almost identically on {the} CPU} {as measured by wall clock time}. {Furthermore,  we also have} access to the gradient through AD, {with {at least double the computational cost}}.

\section{Accelerated Inference with \synax}
\label{sec:inference}
In this section, we demonstrate the performance of \synax\ when performing inference using gradient-based algorithms on two test cases. In the first example, we use the No-U-Turn sampler (NUTS; \citealt{NUTS}), a variant of Hamiltonian Monte Carlo (HMC; see e.g.\ \citealt{HMC}) to obtain the posterior distribution {of magnetic field model parameters}. The sampler is implemented in \texttt{Blackjax}\footnote{\url{https://blackjax-devs.github.io/blackjax/index.html}} \citep{cabezas2024blackjax}. In the second example, we use gradient-based optimization on extremely high-dimensional 3D grids, {to infer the full magnetic field based on mock observations}.

\subsection{Model Setup}
\label{sec:model}
\begin{figure*}
    \centering
    \includegraphics[width=0.85\linewidth]{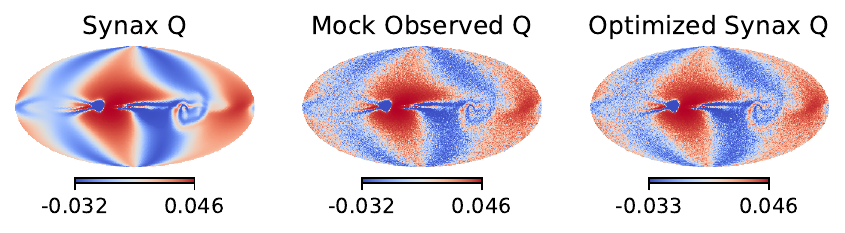}
    \caption{From left to right: simulated synchrotron $Q$ map, mock observation, and optimized synchrotron $Q$ map in units of Kelvin with WMAP $\mathbf{B}$, $N_0$ models and YMW16 $n_e$ model at 2.4 GHz. The map has NSIDE = 64 and $N_{\rm int}=512$.}
    \label{fig:sim_wmap}
\end{figure*}
In our mock observation, we simulate the $\{Q,U\}$ maps with \synax. The $N_0$ model here is the one adopted by WMAP \citep{Drimmel2001,Page2007},
\begin{equation}
    N_0(\mathbf{r}) = C_{0}e^{-\rho/h_r}{\rm sech}^2(z/h_z),
\end{equation}
where $\mathbf{r}=(x,y,z)$ and $\rho = (x^2+y^2)^{1/2}$. We fix the free parameters $h_r = 5 \,{\rm kpc}$ and $h_z = 1 \,{\rm kpc}$, {which are} the original WMAP parameter values. The value of $C_0$ is fixed by $N_{0,\rm Earth} = 4.0\times 10^{-5} {\rm cm}^{-3}$, consistent with the observations of 10 GeV electrons on Earth \citep[e.g.][]{Strong2007}. The $n_e$ model is the YMW16 model \citep{YMW16}, a complex model {that accounts for} disks, arms, Galactic Loop I, and the Local Bubble. The $\mathbf{B}$ model is also the WMAP model \citep{Page2007},
\begin{equation}
\begin{aligned}
    \mathbf{B}(\rho,\phi,z) &= b_0[ \cos(\psi(\rho))\cos(\chi_B(z))\hat{\rho}\\
    &+\sin(\psi(\rho))\cos(\chi_B(z))\hat{\phi}+ \sin(\chi_B(z))\hat{z}],
\end{aligned}
\end{equation}
where $\{\rho,\phi,z\}$ are the cylindrical coordinates centered at the Galactic center and $\{\hat{\rho},\hat{\phi},\hat{z}\}$ are the corresponding unit vectors. For $\rho\in[3,20]$ kpc, $\psi(\rho) = \psi_0 +\psi_1\ln(\rho/8{\rm kpc})$ and $\chi_B(z)=\chi_{0,B}\tanh(z/1{\rm kpc})$. $\psi(\rho)$ and $\chi_B(z)$ are both zero otherwise. The true values of the free parameters in the model are set to $\{b_0,\psi_0,\psi_1,\chi_{0,B}\} = \{1.2\times10^{-6}\ {\rm gauss, 0.4712\ {\rm rad},0.0157\ {\rm rad},0.4363\ {\rm rad}}\}$. We simulate the maps with NSIDE = 64 and $N_{\rm int}=512$, corresponding to a resolution from {10 to 56 pc} along different LOS. The input for $\mathbf{B}$ and $N_0$ are callable functions and the $n_e$ input is a 3D {regular} grid with resolution $(256,256,64)$. The simulated $Q$ map is shown in the left panel of Figure \ref{fig:sim_wmap}. A mock observation is then created by adding {independent} Gaussian noise with noise standard deviation $\sigma_n=1$ mK, as is shown in the middle panel of Figure \ref{fig:sim_wmap}.

We then use \synax\ to infer $\mathbf{B}$. The $n_e$ and $N_0$ inputs are identical to the simulation, and we keep $N_{\rm int}=512$ to avoid numerical errors from insufficient resolution. We set the {priors for the magnetic field parameters} as 
\begin{equation}
\begin{aligned}
    b_0 \sim \mathcal{U}[0,10],\\
    \psi_0 \sim \mathcal{U}[0,\pi/2],\\
    \psi_1 \sim \mathcal{U}[0,\pi/2],\\
    \chi_{B,0} \sim \mathcal{U}[0,\pi/2],
\end{aligned}
\end{equation}
where $\mathcal{U}[a,b]$ represents a uniform prior from $a$ to $b$. The noise in the simulated observations is Gaussian and independent between pixels. The log-likelihood for this model therefore corresponds to the Gaussian likelihood with standard deviation $\sigma=1$ mK for each pixel,
\begin{equation}
    \log \mathcal{L} = {-}\sum_i \left(\frac{(Q_i - Q_{i,\rm obs})^2}{2\sigma^2} + \frac{(U_i - U_{i,\rm obs})^2}{2\sigma^2}\right) + C
\end{equation}
where $C$ is a normalization constant. 

\subsection{Sampling with NUTS}

\begin{figure}
    \centering
    \includegraphics[width=\linewidth]{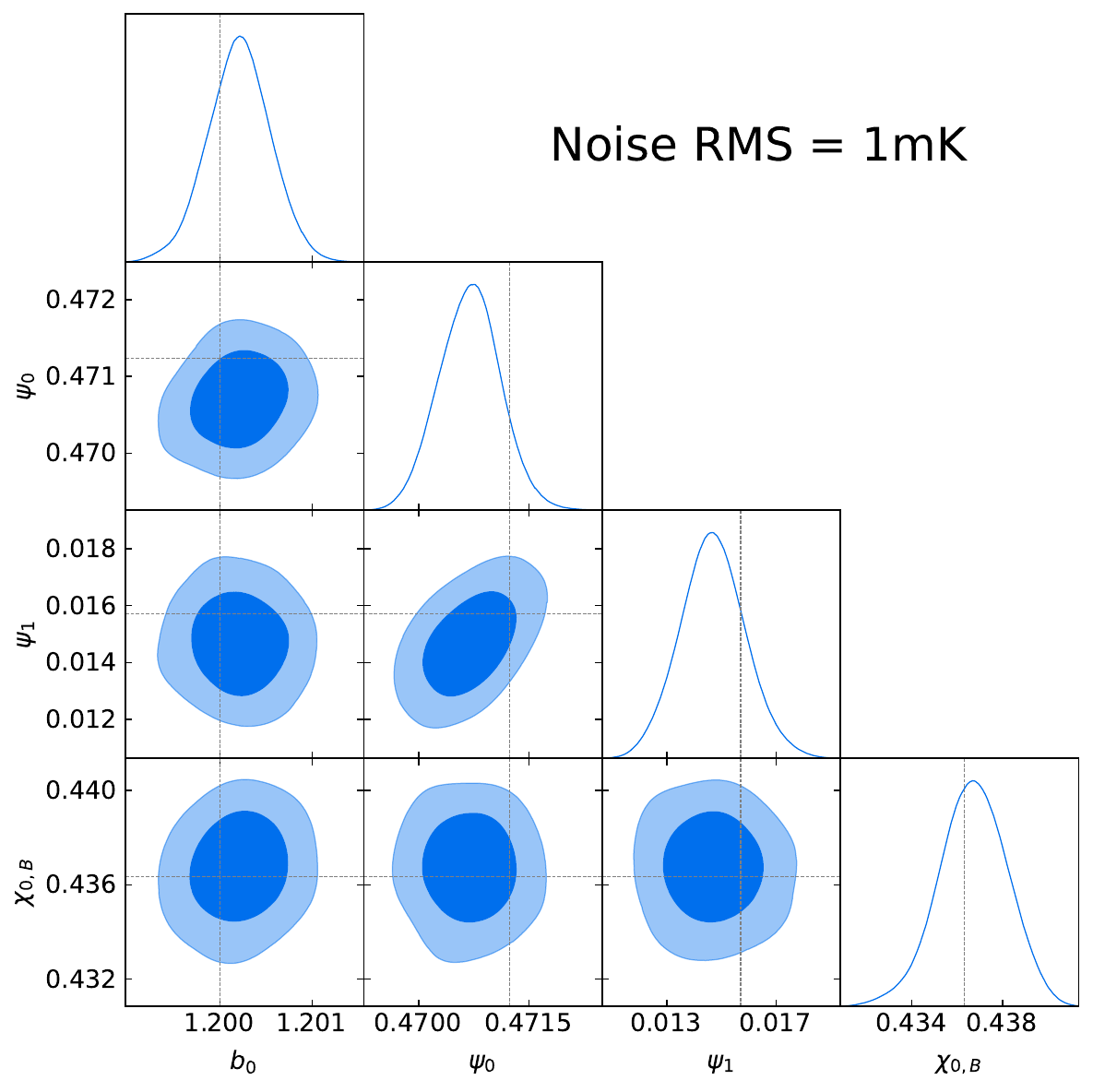}
    \caption{Posteriors for WMAP $\mathbf{B}$ field parameters obtained with NUTS. The shaded areas are the $1\sigma$ (dark blue) and $2\sigma$ (light blue) credible regions, and gray dashed lines indicate the true parameter values. The mock observation map consists of a simulated noiseless map and white noise with standard deviation of 1 mK.}
    \label{fig:post}
\end{figure}
\begin{table}
    \centering
    \caption{Posterior summary statistics and diagnostics obtained with NUTS.}
    \begin{tabular}{lcccc}
    \hline
         Parameter&Best-fit & Accuracy& $\hat{r}$&ESS \\\hline
         $b_0$[$\times10^{-6}$gauss]& $1.2002\pm 0.0003 $&0.01\% &1.003 & 553.35\\
         $\psi_0$[rad]& $0.4707\pm 0.0004$&-0.11\% &1.003 & 503.94\\
         $\psi_1$[rad]& $0.0147\pm0.0012$&-6.48\% &0.999 & 310.59\\
         $\chi_{0,B}$[rad]& $0.4367\pm 0.0014$&0.09\% &1.006 & 252.37\\
         \hline
    \end{tabular}
    \label{tab:nonoise}
\end{table}
\begin{table}
    \centering
    \caption{ESS per second obtained with NUTS and RWMH after burn-in.}
    \begin{tabular}{lcccc}
    \hline
         Method&$b_0$ & $\psi_0$&$\psi_1$&$\chi_{0,B}$ \\\hline
         NUTS GPU& 1.90 &1.73&1.07 & 0.87\\
         NUTS CPU& 0.09 &0.06&0.04 & 0.03\\
         RWMH GPU& 9.49 &13.30&7.34 & 0.37\\
         RWMH CPU& 0.14& 0.20&0.11& 0.01\\
         {RWMH CPU \texttt{hammurabi}}& {0.03}& {0.005} & {0.006} & {0.003} \\ 
         \hline
    \end{tabular}
    \label{tab:ess}
\end{table}
We first sample the posterior of the WMAP analytical $\mathbf{B}$ field model parameters, i.e. $\{b_0,\psi_0,\psi_1,\chi_{0,B}\}$, with NUTS, {with other parameters and settings fixed to those described} in Section \ref{sec:model}.  We run NUTS for 600 iterations and leave the first 100 as burn-in. {A single forward model and gradient} evaluation requires $\sim$100 ms to finish, and such a {full sampling} run takes $\sim $ 7 minutes on an NVIDIA Tesla A100 GPU. The posterior is shown in Figure~\ref{fig:post}. The {Gelman}-Rubin statistic  $\hat{r}$ \citep{rhat} is a convergence diagnostic used in computational Bayesian statistics to assess the mixing and convergence of MCMC chains. It compares the variance within and between multiple chains, {with values $\hat{r}<1.01$ being indicative of convergence. One can also calculate the effective sample size (ESS), which provides an estimate for the number of independent samples in the correlated chains. For the example in this section, the values of $\hat{r}$ are all below 1.01, and the ESS is of the same order as the number of sampling iterations.} We can therefore be confident that we have converged on the stationary distribution, and the corresponding posterior samples are generated with minimal auto-correlation. {In Table \ref{tab:nonoise} we provide the best-fit posterior expectations for each parameter, the corresponding accuracy (defined as $(\Bar{p} - p_{\rm true})/p_{\rm true}$, for parameter expectation $\Bar{p}$ and true value $p_{\rm true}$), alongside the $\hat{r}$ and ESS diagnostics for each parameter. The accuracy is close to zero, except for $\psi_1$ due its small magnitude. Nonetheless, the true parameter value lies within the 1$\sigma$ credible interval.}

We also run NUTS on {the} CPU and Random Walk Metropolis-Hastings (RWMH) on the GPU and CPU to provide a benchmark, with the ESS per second for each parameter being reported in Table \ref{tab:ess}. {We use \synax\ to run simulations on the CPU, given that it has similar performance to \texttt{hammurabi}.} On {the} CPU, a single \synax\ realization with 128 threads takes $\sim 200$ ms to finish. {To ensure the reliable computation of posterior expectations from MCMC samples, one must obtain a sufficient number of effective samples. The ESS per second therefore provides a measure of the computational efficiency of each algorithm. Comparing the minimum ESS per second (i.e. ESS per second for $\chi_{0,B}$), we obtain approximately a 20 times speed-up from GPU acceleration, with NUTS providing an additional factor of two improvement for this simple example model.} 
The improvements from using gradient-based sampling methods such as NUTS will be more apparent for higher dimensional sampling problems \citep{gelman1997weak,beskos2013hmc}. {We further compare the performance of \texttt{hammurabi} on the same problem, using RWMH for inference.}\footnote{{We construct the mock observation with \texttt{hammurabi} to avoid numerical discrepancies between different simulations. Specifically, we use the same default physical parameters and adjust the integration hyper-parameters such that the total number of integration points remains approximately the same as in \synax.}} 
{The results are shown in the bottom row of Table \ref{tab:ess}. Using \texttt{hammurabi} we also reach a converged posterior identical to that obtained with \synax, albeit with a lower ESS per second. This performance degradation can mainly be attributed to Python’s lower efficiency. In \synax, we bypass this issue by compiling the entire pipeline with JIT compilation provided by JAX, as both the simulator and the inference pipeline is built within the JAX ecosystem.}


\subsection{Optimizing 3D Grids}
\label{sec:optlsa}
\begin{figure}
    \centering
    \includegraphics[width=\linewidth]{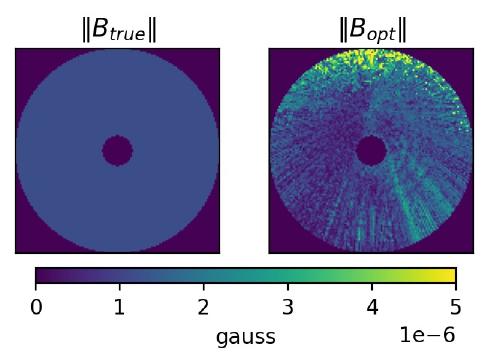}
    \caption{Left: a slice of the true $\mathbf{B}$ field magnitude at z = 0 kpc. Right: a slice of the optimized $\mathbf{B}$ field magnitude at z = 0 kpc.}
    \label{fig:optb}
\end{figure}

The WMAP analytical modeling of the GMF focuses on the large-scale structure of the Galaxy, but does not account for small-scale features such as turbulence in the Galactic plane. In this section, we consider a more general approach by directly optimizing the 3D $\mathbf{B}$ grids based on mock observations. The grid consists of $128\times128\times32$ voxels, with a corresponding box size of $(40,40,10)$ kpc. Our goal is to optimize the 3D grids to maximize the log-likelihood $\log \mathcal{L}$. Due to the high dimensionality ($\sim 5\times 10^6$) of this problem, we employ the ADAM optimizer \citep{ADAM} implemented {in} \texttt{Optax} \citep{deepmind2020jax}\footnote{\url{https://optax.readthedocs.io/en/latest/}}. Optimization is halted after 200 iterations. {Running the optimization further results in the loss function oscillating around the minimum, indicating the optimization has converged.}

The optimized synchrotron map is presented in the right panel of Figure \ref{fig:sim_wmap}. Due to the high flexibility of the grid representation, noise features are also reproduced, indicating significant overfitting. The optimized $\mathbf{B}$ magnitude is shown in Figure \ref{fig:optb}, where it is evident that the $\mathbf{B}$ field fluctuates considerably to replicate the noise features. Although this result is not physically valid, the striking similarity between the optimized maps and the mock observations demonstrates the model's ability to reproduce observations. This also suggests that regularization is required to effectively prevent overfitting when dealing with fields possessing many degrees of freedom.

This optimization would be intractable without AD. Traditional methods such as finite difference, exemplified by MIUNIT \citep{1975CoPhC..10..343J}, are often used for optimization tasks \citep[e.g.][]{2024ApJ...970...95U}. However, these methods generally require $\mathcal{O}(n)$ calls to compute the numerical gradient, where $n$ represents the degrees of freedom. In our field-level optimization problem, $n = 128\times128\times32$, making the computation of the numerical gradient intractable.

\section{Discussion}
\label{sec:discussion}
\subsection{Systematics Analysis}
\begin{figure}
    \centering
    \includegraphics[width=\linewidth]{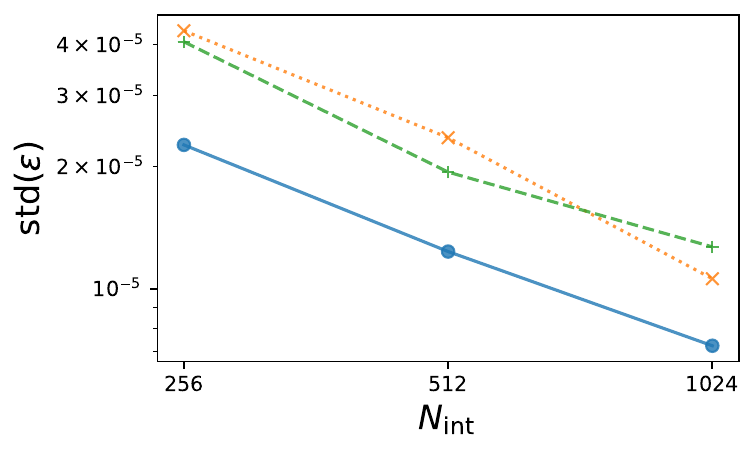}\\
    \includegraphics[width=0.98\linewidth]{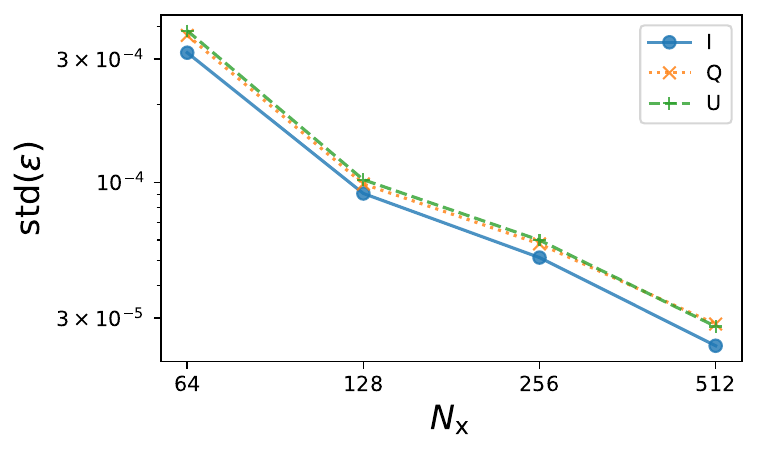}
    \caption{\textbf{Upper}: Standard deviation of the residuals for the simulated maps with different $N_{\rm int}$, defined with respect to the map generated with $N_{\rm int}=2048$. \textbf{Lower}: Standard deviation of the residuals of a grid $\mathbf{B}$ model map compared to an analytical $\mathbf{B}$ model map, as a function of grid size $N_x$. Solid, dotted, and dashed lines represent $\{I,Q,U\}$ maps respectively.}
    \label{fig:err}
\end{figure}

In this section, we analyze the errors introduced by numerical integration and grid interpolation. We fix NSIDE = 64 throughout this subsection. First, we study the integration error obtained by varying the number of integration points, {$N_{\rm int} \in \{256, 512, 1024, 2048\}$}, which corresponds to a maximum resolution ranging from {130 to 20 pc}. All other hyper-parameters and physical models are identical to those described in Section \ref{sec:model}. The systematic error is evaluated by calculating the standard deviation of the residuals relative to the $N_{\rm int} = 2048$ map, defined as ${\rm std}(\epsilon) = {\rm std}({\rm Map}_{N_{\rm int}} - {\rm Map}_{2048})$. The variation of ${\rm std}(\epsilon)$ with $N_{\rm int}$ is shown in the upper panel of Figure \ref{fig:err}. We find that the typical scale of systematic errors is less than $4\times 10^{-5}$ K, even with $N_{\rm int} = 256$, which is significantly below the 1 mK sensitivity of the 2.3 GHz S-PASS survey \citep[e.g.][]{S-PASS}, which we treat as typical for the frequency range we consider. Consequently, it is safe to use a relatively small value of $N_{\rm int}$ to expedite the inference process, {with the numerical integration errors being safe to ignore} when evaluating the likelihood.

Secondly, we assess the error associated with using 3D grid and interpolation functions as field generators. The fiducial model in this analysis is identical to that described in Section \ref{sec:model}. To evaluate the interpolation error, we first generate a $\mathbf{B}$ field on regular grids of size $(N_x, N_x, N_x/4)$ using the WMAP analytical function, and then interpolate these fields onto the desired integration points to calculate the resulting synchrotron emission. We choose $N_x$ {$\in \{64, 128, 256, 512\}$}. The systematic error is characterized by ${\rm std}(\epsilon) = {\rm std}({\rm Map}_{N_{x}} - {\rm Map}_{\rm ana})$, where ${\rm Map}_{\rm ana}$ represents the synchrotron map generated using the analytical expression, which serves as our fiducial model, and $N_{\rm int}$ is fixed at 512. The results are shown in the lower panel of Figure \ref{fig:err}. We observe that with $N_x = 64$, the error reaches sub-mK levels, making it non-negligible during the inference process. {As the grid size increases, the error level decreases, with ${\rm std}(\epsilon)\sim10^{-5}$ K when $N_x = 512$, indicating an insignificant error contribution from interpolation at this resolution}. However, it is important to note that the fiducial model here assumes a smooth $\mathbf{B}$ field, which reduces the interpolation error. A higher level of systematic error is possible for a turbulent $\mathbf{B}$ field with low smoothness, especially when using smaller $N_x$ values.

\subsection{Detailed Performance profiling}
\begin{figure}
    \centering
    \includegraphics[width=\linewidth]{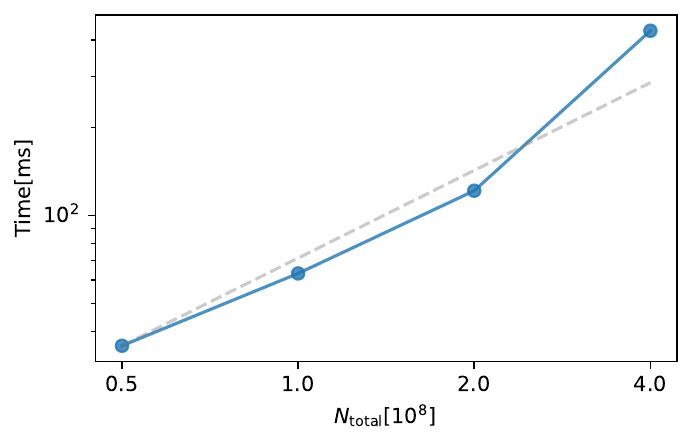}
    \caption{The blue solid line indicates wall clock time for a single realization with respect to the total number of integration points $N_{\rm total}$, using a Tesla A100 GPU. The grey dotted line corresponds to the theoretical scaling where time is proportional to $N_{\rm total}$. }
    \label{fig:time}
\end{figure}
{In this section we analyze the time elapsed in performing one simulation for different values of the total number of integration points $N_{\rm total}$. We can evaluate $N_{\rm total}$ using $N_{\rm total} = 12\times{\rm NSIDE}^2\times N_{\rm int}$, with NSIDE = 128.} Ideally, the runtime should be proportional to $N_{\rm total}$, since the operations {for a given} integration point have constant computational complexity. We change $N_{\rm total}$ by setting different {$N_{\rm int} \in \{256,512,1024,2048\}$}, with the corresponding runtime being shown in Figure \ref{fig:time}. We find that the actual runtime has a similar trend to the theoretical scaling, indicating the workload is equally distributed across GPU cores, and the GPU utilization is nearly optimal. 

For $N_{\rm total}>10^8$, the runtime of a single realization exceeds $\sim50$ ms. {Evaluating the gradient is four times more expensive in runtime {in} our test, with negligible increase as the number of field parameters grows}. {However, this} additional computational cost is justified by the ability to employ gradient-based inference algorithms, which scale more effectively with dimensionality, and can be designed to fully leverage the parallel computation capabilities of GPUs and TPUs \citep{Hoffman2021AnAS}. 

\subsection{Sampling with a 16-parameter model}

\begin{table}
    \centering
    \caption{Model parameters for four WMAP \textbf{B} field models in the 16-parameter model analysis.}
    \begin{tabular}{lcccc}
    \hline
         Model&$b_0$[$\times 10^{-6}$gauss] & $\psi_0$[rad]&$\psi_1$[rad]&$\chi_{0,B}$[rad] \\\hline
         Model 1& 1.20 &0.4712&0.0157 & 0.4363\\
         Model 2& 3.50 &0.6457&0.5393 & 0.6108\\
         Model 3& 7.70 &1.1693&1.0629 & 1.1344\\
         Model 4& 5.70 & 1.6930&1.5865& 1.6580\\
         \hline
    \end{tabular}
    \label{tab:mod}
\end{table}




To assess the performance of \synax\ with gradient-based sampling methods in higher dimensions, we construct a mock observation using a 16-parameter model. This model is based on four WMAP magnetic field configurations, denoted as $\{\mathbf{B}_1,\mathbf{B}_2,\mathbf{B}_3,\mathbf{B}_4\}$, each characterized by different parameter sets shown in Table \ref{tab:mod}. The total magnetic field, $\mathbf{B}_{\rm total}$, is generated by stacking these components with randomly chosen weights,  
\begin{equation}
\begin{aligned}
    \mathbf{B}_{\rm total} &= (0.6B_{1,x}+0.3B_{2,x}+0.1B_{3,z}+0.3B_{4,y})\hat{x}\\
    &+(0.6B_{1,y}+0.3B_{2,z}+0.1B_{3,y}+0.3B_{4,x})\hat{y}\\
   & +(0.6B_{1,z}+0.3B_{2,y}+0.1B_{3,x}+0.3B_{4,z})\hat{z},
\end{aligned}
\end{equation}
where $\{B_{*,x},B_{*,y},B_{*,z}\}$ represent the three components of the magnetic field, and $\{\hat{x},\hat{y},\hat{z}\}$ are unit vectors along the respective axes. To avoid multimodality in the posterior distribution, $\{\mathbf{B}_2,\mathbf{B}_3,\mathbf{B}_4\}$ are rotated before stacking. Other fields and a noisy mock observation are then generated using the same model described in Section \ref{sec:inference}.

We compare the performance of the NUTS and RWMH algorithms on {the} GPU for sampling this more complex model. NUTS achieves a minimum ESS per second of $1.53 \times 10^{-2}$, whereas RWMH fails to converge within 12 hours, demonstrating the clear advantage of NUTS for efficiently exploring more high-dimensional parameter spaces.

\subsection{Reproducing the Haslam Map}
\begin{figure*}
    \centering
    \includegraphics[width=\linewidth]{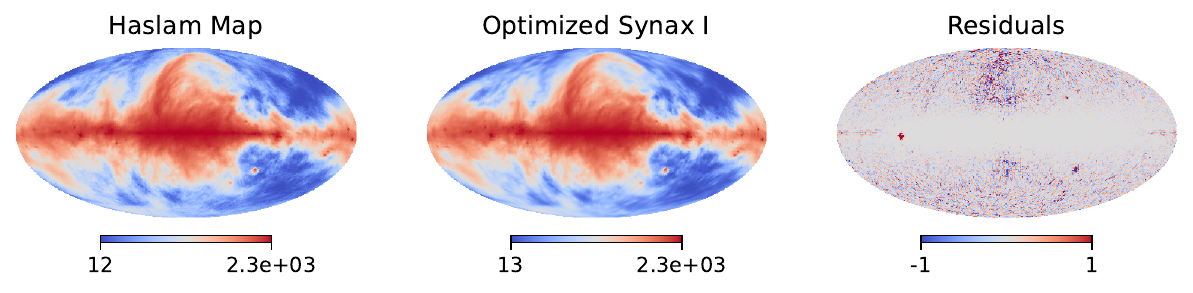}
    \includegraphics[width=0.7\linewidth]{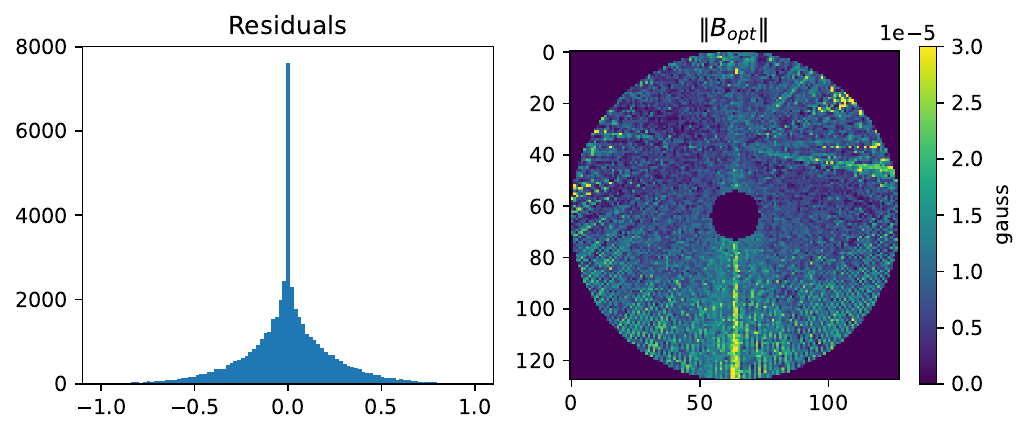}
    \caption{Reproduced Haslam map by optimising the $\mathbf{B}$ field on grids in units of Kelvin. \textit{Top left}: Haslam map spatially averaged to NSIDE = 64. \textit{Top middle}: Synchrotron $I$ map generated with optimized $\mathbf{B}$ field. \textit{Top right}: residual map of the optimised $I$ map compared with Haslam map. \textit{Bottom left}: histogram of residuals. \textit{Bottom right}: a slice of optimised $\mathbf{B}$ magnitude at z=0 kpc.}
    \label{fig:haslam-map}
\end{figure*}

To test the performance of our code in a more complex and realistic situation, we optimize the 3D $\mathbf{B}$ field on grids to generate a simulation based on the Haslam map\footnote{\url{https://lambda.gsfc.nasa.gov/product/foreground/fg_2014_haslam_408_get.html}} \citep{Haslam}.
We downgrade the map to NSIDE = 64 by spatial averaging and neglect beam effects since the pixel size is approximately equivalent to the Haslam beam width, FWHM = $56\ {\rm arcmin}$.
We use identical settings to Section \ref{sec:optlsa} {and again} maximize the likelihood. The results are shown in Figure \ref{fig:haslam-map}. The optimized $\mathbf{B}$ field faithfully reproduces the Haslam map, with a residual standard deviation of $0.29$ K. We see from the residual map that, except for some bright spots that are likely to be localized objects, the reconstruction error is very close to zero. However, the optimized $\Vert \mathbf{B}\Vert$ shows radial features, indicating strong degeneracy along the LOS direction. We expect additional data in intensity and polarization {with more frequency channels} to break the degeneracy in future work, {and data-driven priors learned from magneto-hydrodynamical simulations {also have the potential to help} regularize the results \citep[e.g.][]{2024ApJ...975..201F}}. 

\section{Conclusion}
\label{sec:conclusion}
{As future radio observations targeting cosmological probes such as 21~cm {intensity mapping} and CMB polarization aim at increasingly precise measurements, fast and accurate generative modeling of Galactic synchrotron emission will be vital for inference tasks, both for understanding Galactic structure and evolution, and in mitigating foreground contamination in cosmological signals. However, higher-fidelity models of Galactic synchrotron emission come at an increased cost in computation and dimensionality, demanding the use of gradient-based algorithms and hardware acceleration for tractable inference.}

In this paper, we introduce \synax, a novel synchrotron intensity and polarization simulation package that leverages AD and hardware acceleration for the first time. GPU acceleration significantly enhances computational speed, while AD enables efficient, gradient-based sampling algorithms for Bayesian posterior inference. Using the full 3D representation of fields with AD-based optimization algorithms allows for the characterization of more complex Galactic field structures. \synax\ facilitates rapid inference with complex models, contributing to a more precise understanding of the Galaxy.

We validate the accuracy {and performance} of \synax, {starting from a simple homogeneous test case}, demonstrating that the error budget remains below 1 mK in this case. We conduct a detailed analysis of accuracy, beginning with errors arising from insufficient integration step sizes. We find that the standard deviation of the error remains well below the milli-Kelvin level for maximum step sizes under {130 pc}. Similarly, interpolation errors with 3D grid representations maintain similar accuracy for grid resolutions below {80 pc}. {In terms of performance, \synax\ can produce a map with $5\times10^{7}-5\times10^8$ integration points in approximately $16-300$ ms on an NVIDIA A100 GPU, which is more than 40 times faster than \synax\ on the CPU, while CPU \synax\ is almost identical to \texttt{hammurabi}.}

Utilizing a mock observation based on the four-parameter WMAP GMF model, we evaluate the efficiency of NUTS in obtaining parameter posteriors. Our results show a twenty-fold increase in computational efficiency due to GPU acceleration and an additional two-fold improvement in sampling efficiency with NUTS compared to the RWMH algorithm as applied a four-parameter model, as measured by the ESS per second. On a more complex 16-parameter model, RWMH does not converge, while NUTS with \synax\ obtains an ESS per second of $1.53\times10^{-2}$. Furthermore, we test the use of 3D grids for field representations, finding that while the 3D field reproduces the mock observation with high fidelity, {it significantly overfits noise features}, indicating the need for further regularization.


In this work, we neglect secondary effects such as self-absorption and conversion \citep[e.g.][]{Jones1977,2019MNRAS.484.1427C} which will be addressed in future updates. Additionally, while free-free emission and absorption, which are negligible beyond the GHz range, are not considered here, they play a critical role in low-frequency observations, which are essential for probing the early Universe. Future iterations of \synax\ will include these processes to extend its applicability to the low-frequency domain.
\begin{acknowledgments}
KND, RDPG and YM are supported by the National SKA Program of China (grant No.~2020SKA0110401) and NSFC (grant No.~11821303). This work is also supported by U.S. Department of Energy, Office of Science, Office of Advanced Scientific Computing Research under Contract No. DE-AC02-05CH11231 at Lawrence Berkeley National Laboratory to enable research for Data-intensive Machine Learning and Analysis.
\end{acknowledgments}

%

\vspace{5mm}


\software{JAX \citep{deepmind2020jax}, Optax, blackjax \citep{cabezas2024blackjax}, \hpy\ \citep{Zonca2019,2005ApJ...622..759G}, \texttt{hammurabi} \citep{hammurabi,hammurabiX}, \texttt{numpyro} \citep{phan2019composable,bingham2019pyro}.
          }




\bibliography{example}{}
\bibliographystyle{aasjournal}



\end{document}